\documentclass[%
 aip,
 cp,  
 amsmath,amssymb,%
 reprint,%
 author-year,%
 numerical
]{revtex4-2}

\usepackage{graphicx}
\usepackage{dcolumn}
\usepackage{bm}

\usepackage[utf8]{inputenc}
\usepackage[T1]{fontenc}
\usepackage{newtxtext}
\usepackage{xcolor}
\usepackage[breaklinks,colorlinks,citecolor=darkgray,linkcolor=black,urlcolor=darkgray]{hyperref}

\begin{document}

\title{Managing Work-Life Balance During the COVID-19 Crisis:\protect\newline
  A survey with more than 1,500 participants}

\author{Ruzin A\u{g}ano\u{g}lu}
 \email{akc@dpg-physik.de}
\affiliation{%
  Working Group on Equal Opportunities of the German Physical
  Society, Hauptstra{\ss}e 5, 53604 Bad Honnef, Germany}
\affiliation{AGANER Holding, Pommernstra{\ss}e 32, 96242 Sonnefeld, Germany}

\author{Beate Kl\"{o}sgen}
 \email{akc@dpg-physik.de}
\affiliation{%
  Working Group on Equal Opportunities of the German Physical
  Society, Hauptstra{\ss}e 5, 53604 Bad Honnef, Germany}
\affiliation{Department for Physics, Chemistry, and Pharmacy, University of
  Southern Denmark, Campusvej 55, 5230 Odense M, Denmark}

\author{Agnes Sandner}
 \email{akc@dpg-physik.de}
\affiliation{%
  Working Group on Equal Opportunities of the German Physical
  Society, Hauptstra{\ss}e 5, 53604 Bad Honnef, Germany}

\author{Iris Traulsen}
 \email[Corresponding author: ]{itraulsen@aip.de, akc@dpg-physik.de}
\affiliation{%
  Working Group on Equal Opportunities of the German Physical
  Society, Hauptstra{\ss}e 5, 53604 Bad Honnef, Germany}
\affiliation{Leibniz-Institut f\"{u}r Astrophysik Potsdam (AIP), An der
  Sternwarte 16, 14482 Potsdam, Germany}

\date{\today}

\begin{abstract}
  In the early phase of the COVID-19 pandemic, the Working Group on Equal
  Opportunities of the German Physical Society conducted a nonrepresentative,
  international online survey, investigating the impact of the pandemic and
  its consequences on daily life and work. Special attention was paid to the
  situation of female physicists in academia and industry. The responses
  provided evidence of both difficulties and opportunities arising from
  adapting to the rapid, significant changes. Discrepancies between different
  countries, jobs, and gender were found, particularly regarding concerns
  about professional work and private life conditions after the pandemic.
\end{abstract}

\maketitle

\section{\label{sec:context}Introduction: Context of the COVID-19 survey}

The global spread of COVID-19 forced governments worldwide to implement
protective measures within a short time frame. In a rare coincidence,
employees all over the world were facing similar challenges and similar
changes to their daily routines, including contact reductions, closure of
schools and workplaces, and trading and travel restrictions. According to the
Committee for the Coordination of Statistical Activities of the United
Nations, the pandemic caused the largest global economic and employment
decline since 1945 \cite{unesco_uis}. Within its first weeks, most countries
closed schools, shifting education and caring obligations to the families of
about 1.6 billion students and imposing gendered impacts on both students and
carers \cite{unesco_education}. While similar actions were taken more or less
all over the world, the social and economic consequences and the potential
impacts of all aspects of the pandemic varied between countries and between
rural regions and cities \cite{oecd}.

From April 2020 onward, the Working Group on Equal Opportunities of the German
Physical Society investigated living and working conditions, organizational
challenges, and concerns about the postpandemic consequences in an online
survey. The focus of the survey was on European women in physics, both in and
outside academia, but it was directed to workers in other disciplines as
well. Major results are presented and discussed in the following sections.

\section{\label{sec:methods}Conducting and evaluating the survey}

The survey was conducted fully anonymously and online in 2020 from mid-April
until the end of June. Twelve national and international nonprofit societies
in physics, chemistry, engineering, innovation, and economics spread it,
particularly inviting women working in physics and other fields related to
science, technology, engineering, and mathematics (STEM) to participate. It
consisted of 27 multiple-choice questions in five categories, some with the
option to add a comment, and one free-text area. The questions addressed (1)
the education, work, and living situation of the participant; (2) the impact
of the pandemic on private life; (3) the impact of the pandemic on
professional life, (4) the participant’s experience level regarding time
management and leadership; and (5) expectations and concerns regarding
postpandemic conditions.

The analysis of the responses was based on all completely submitted
questionnaires, while partly answered forms were excluded. The survey was not
designed to be representative, and therefore statistical significance was not
calculated for the responses. Three questions that allowed for comments were
processed using topic modeling.  We employed a Latent Dirichlet Allocation
model \cite{lda} to identify the major topic groups of the answers and their
prevalence. The responses were evaluated separately according to gender, age
group, education, employment in or outside academia, leadership position, and
continent of residence.

\section{\label{sec:results}Survey results}

Out of 2,209 participants who started to answer the questionnaire, 1,524
belonged to the target group of people of working age and completed the
questionnaire. Reflecting the channels by which the survey was announced, a
majority (65\%) of the forms were answered by women, 43\% by individuals
with a background in physics, and 29\% by individuals with a background in
other STEM disciplines. Eighty-four percent of the participants lived in
Europe at the time of the survey, 8\% in North America, 5\% in Asia, and
3\% in South America and Africa. Seventy percent worked in academia, and
32\% held a leadership position.

Most participants (82\%) reported relevant changes to their private life due
to the pandemic, and almost all participants reported changes to their working
conditions. Detailed information about the participants and responses is
provided online by the Working Group on Equal Opportunities of the German
Physical Society \cite{akcpdf}, and a summary was published in German
\cite{pj}.

  \subsection{\label{sec:questions}Questionnaire Results and the Disparities between Groups}

\begin{figure}
    \centering
    \includegraphics[width=\linewidth]{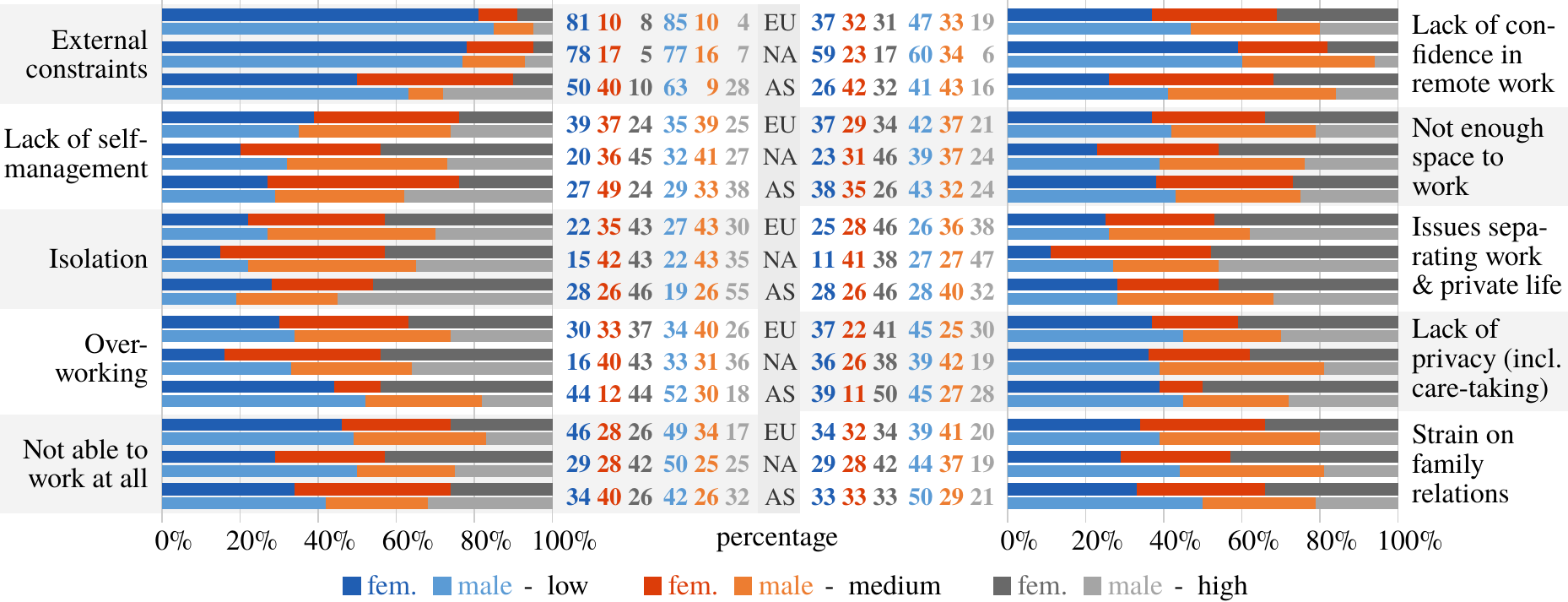}
    \caption{\label{fig:stressfacs}\centering Work-related stress factors in
      the early phase of the pandemic for the three continents with the
      highest response rates: Europe (EU, 888 women, 389 men), North America
      (NA, 53 women, 75 men), and Asia (AS, 24 women, 53 men). In the paired
      bars, top bar (darker colors) is answers by women (w), bottom bar
      (lighter colors) is answers by men (m). The corresponding percentages
      are provided using the same color coding as their bar representation.}
  \end{figure}

  In all age, gender, and regional groups, the majority of participants
  assigned medium to high stress levels to the changes they experienced in
  their professional and private life. Figure~\ref{fig:stressfacs} illustrates
  the feedback on several work-related questions by female and male
  participants in Europe, North America, and Asia. They considered living and
  working in isolation and issues in separating private from professional life
  as the highest stress factors. The majority of European and North American
  participants saw an increased risk of overworking. Women were more likely
  than men to have experienced a lack of privacy or to have had to cease
  working completely due to family obligations.

\begin{figure}
    \centering
    \includegraphics[width=\linewidth]{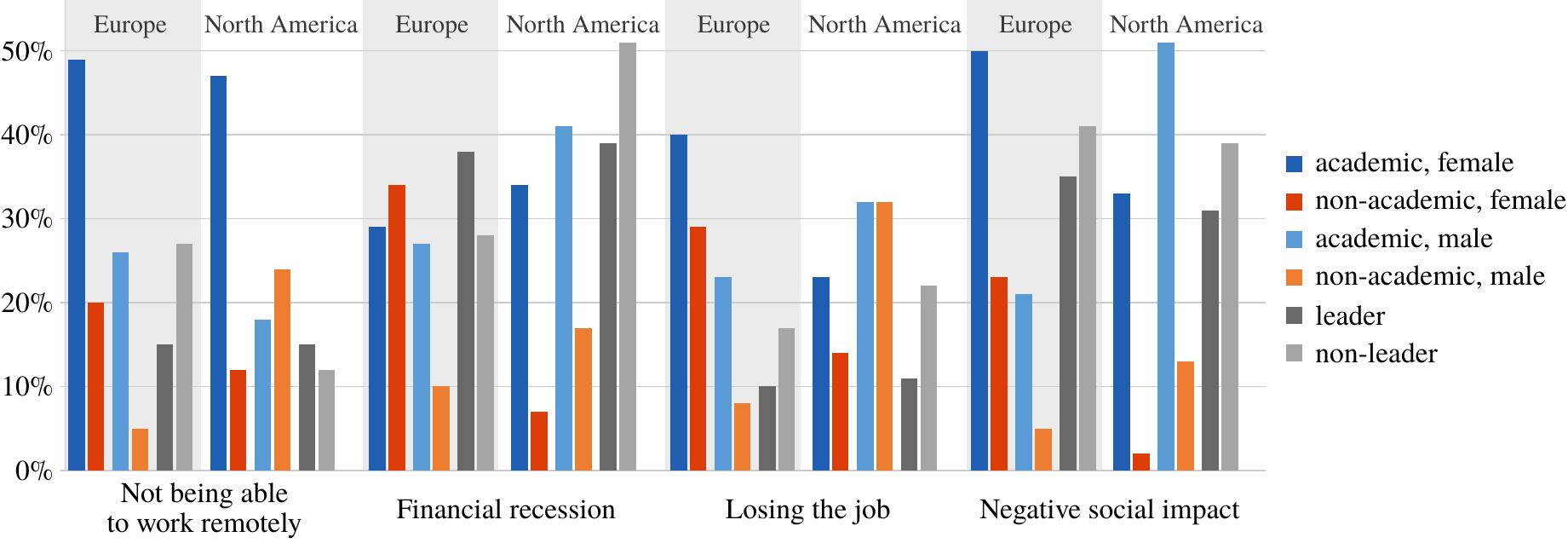}
    \caption{\label{fig:concerns}\centering Postpandemic concerns of women and
      men with academic and nonacademic backgrounds as well as of leaders and
      nonleaders in Europe and North America, the continents with the two
      highest response rates.}
  \end{figure}

  Individuals with leadership responsibilities reported a higher workload
  while their teams had to work remotely, but they were overall satisfied with
  the performance of their teams under the pandemic conditions. Time
  management skills helped both leaders and employees to cope with the
  changes. Larger discrepancies between the different groups were found with
  respect to concerns regarding the consequences after the pandemic
  (Fig~\ref{fig:concerns}). Women and nonleaders doubted that they would have
  sufficient liberty to work remotely. Concerns about financial regression
  were prevalent among North Americans working in academic jobs. In all
  regions, academics were the most afraid of losing their jobs owing to the
  pandemic, potentially related to the large fraction of nonpermanent
  contracts in academia. Few differences were found here between leaders and
  employees. Large spreads can be seen in social concerns, again with the
  highest shares among people in academia. With respect to educational
  background, people working in physics and other STEM fields were less
  worried about postpandemic issues than non-STEM participants.

  \subsection{\label{sec:freetext}Free-text Answers and the Relevance of Work-Life Balance}

  Missing work-life balance was the predominant topic of the free-text
  answers. The very detailed texts unveiled gender-related roles under
  pandemic conditions. Parents and women in particular reported an increased
  load due to home-schooling children, organizing the family life in
  isolation, and caring for elder relatives and members of risk
  groups. High-risk individuals and those with impairments, as well as their
  contact people, experienced especially hard restrictions. Families with
  preschool and primary school children faced particular difficulties in
  arranging childcare, education, and normal work hours. Single people
  described the effects of isolation both professionally and privately. All
  groups noticed mental and emotional stress and the risk of reduced
  productivity at work.

  In contrast, positive aspects were also recorded: more time for family,
  partnership, and personal interests in a temporarily decelerated daily
  routine; increased awareness, deference, and solidarity; higher work
  productivity thanks to fewer appointments and routine tasks; and chances to
  learn positive lessons from the crisis and react constructively. Although
  some participants were afraid that a larger share of remote communication
  would reduce personal contacts and influence working conditions negatively
  in the long term, others were confident that the experience gained during
  the pandemic would allow for improved and efficient schooling and work
  strategies in the future. Examples of free-text answers are given below.

  \begin{itemize}

  \item ``Unfortunately, we were not prepared for effective home office. The
    biggest distraction were kids, since schools were relying on parents to
    teach the kids at home.''

  \item ``Loving working from home -- getting so much more thinking work done,
    because I'm not interrupted like at work. (Even though I have a 10 and
    12year old at home)''

  \item ``As a woman and mother, I do not see my perspective represented by
    male politicians.''

  \item ``Perception and discussion culture of what people are currently doing
    for caring at home (children or other people in need of care) is
    missing.''

  \item ``Rediscovering slowness is positive for me, even though I lack social
    contacts. No hectic, no or only very few appointments.''

  \item ``Re-traditionalization of care-work in the areas of responsibility of
    female connotations is a serious social regression.''

  \item ``I enjoy working from home because I get to spend more time with my
    husband and have more control over my working hours. I actually feel less
    stressed out now than before the pandemic and I'm getting more sleep. I
    also have fewer interruptions from my colleagues.''

  \item ``Despite being physically away from work and having potentially more
    time to exercise, my overall health deteriorated.''

  \item ``I am disabled and do not have the same level of support as at my
    workplace to fulfil my duties.''

  \item ``Being from a city that is part of China, I am concerned about
    post-pandemic hostility and discrimination that may rise against me and my
    work.''

  \item ``I feel much better and less stressed working from home and with
    lowered interactions than ever before. The context of the pandemic is
    stressing but the work habit it imposed suits me just fine.''

  \item ``The spread between poor and rich and also between educated and
    non-educated will be widened hugely.''

  \item ``Post-pandemic, I think there is a huge benefit in employers offering
    employees the opportunity to work from home if -- and when -- possible to
    better accommodate their work-life balance. It would be disappointing to
    see some benefits this pandemic has implemented into routine and balance
    and then watch it being taken away when the pandemic is over.''

  \end{itemize}

\section{\label{sec:conclusions}Discussion and conclusions}

At the time of the survey in 2020, governments and societies were struggling
with how to deal with the pandemic and how to establish new, temporary, or
permanent routines. The survey responses indicate that imbalances between
private and work life and between gender roles were strengthened in the early
phase of the pandemic. Partly hidden gender-related responsibilities and
attribution of tasks became more obvious. Reflecting on these disparities can
help to overcome them, not only during the current crisis, but
sustainably. The chance exists to raise awareness for a more balanced sharing
of family obligations. Employees now experience more autonomy and
responsibility regarding their tasks but should have access to structured and
regular feedback and performance recording. Leaders have the chance to
implement improved work and information flows in a phase of structural
transition and flexibility. Society and authorities can benefit from the new
experience and tools. Recognition of the relevance of science and fact-based
decisions has increased, and science needs actively contributing scientists,
irrespective of their backgrounds. Despite the issues, the pandemic offers the
opportunity to break fresh ground, professionally and socially, and to foster
and acknowledge the impact of male, female, and minority role models on the
progress of science and society.

\begin{acknowledgments}

  We thank As{\i}m Can Alkan, P{\i}nar Bilge, Jutta Kunz-Drolshagen, Carola
  Meyer, and Leanna M{\"u}ller for their valuable contributions to the
  survey. Financial support by the German Physical Society for participation
  in the 7th IUPAP International Conference on Women in Physics is gratefully
  acknowledged.

\end{acknowledgments}

\bibliography{akc_germany_survey}

\end{document}